\documentclass{article} \usepackage{graphicx,epstopdf} \oddsidemargin
-1.5cm \evensidemargin -1.5cm \topmargin -1.5cm \textwidth 18cm \textheight
22.5cm

\def\circa#1{\,\raise.3ex\hbox{$#1$\kern-.75em\lower1ex\hbox{$\sim$}}\,}

\newcommand{\psl}{p\hspace{-4.1pt}{\scriptstyle /}}
 
  \newcommand{\be}{\begin{equation}}
\newcommand{\ee}{\end{equation}} \newcommand{\ben}{\begin{displaymath}}
\newcommand{\een}{\end{displaymath}} \newcommand{\ba}{\begin{eqnarray}}
\newcommand{\ea}{\end{eqnarray}} \newcommand{\ban}{\begin{eqnarray*}}
\newcommand{\ean}{\end{eqnarray*}} 
 
\newcommand{\eps}{\varepsilon} 
\newcommand{\kt}{\mbox{$k$}_{\perp }} 
\newcommand{\qt}{\mbox{$q$}_{\perp }} 

\newcommand{\g}{\gamma}

\begin{document}
\vspace{1.cm}

{\centering

 {\Large\bf  Anomalous Sudakov Form Factors}

\vspace{1.cm}

{\bf \large Marcello Ciafaloni}

{\it 
 Dipartimento di Fisica, Universit\'a di Firenze and INFN -
 Sezione di Firenze,\\
via Sansone 1, I-50019 Sesto Fiorentino, Firenze (Italy),\\
E-mail: ciafaloni@fi.infn.it
}

\vspace{0.4cm}

{\bf \large Paolo Ciafaloni}

{\it  INFN - Sezione di Lecce, \\Via per Arnesano, I-73100 Lecce, Italy \\
E-mail: paolo.ciafaloni@le.infn.it}
\vspace{0.4cm}

{\bf \large Denis Comelli}

{\it INFN - Sezione di Ferrara, \\Via Saragat 3, I-44100 Ferrara, Italy\\
E-mail: comelli@fe.infn.it}

}

\vspace{0.3cm} 

\begin{abstract}
While radiative corrections of infrared origin normally depress high energy amplitudes (Sudakov form factors), we find that in some cases resummation of leading effects produces exponentials with positive exponents, giving rise to amplitudes that grow indefinitely with energy. The effect happens in broken gauge theories like the electroweak sector of the Standard Model, and is related to the existence of amplitudes that do not respect the gauge symmetry, and that contrary to expectations do not vanish in the very high energy limit, but rather become dominant. As a working example 
we consider a model with two chiral abelian gauge groups $U'(1)\otimes U(1)$
 with large mass splitting $M_{Z'} \gg M_{Z} $, and we compute  leading
 radiative corrections corrections to the decay of the heavy extra ${Z'}$
 boson into light fermions. For proper fermionic charges,
 the chirality breaking magnetic dipole moment, although mass suppressed, becomes the dominant contribution to the $Z'$ width at very high energies.

\end{abstract}

\section{Introduction}
The study of the asymptotic\footnote{by "asymptotic behavior" we mean the
behavior of cross sections for energies much higher than all SM particles
masses. Also the case of energies higher than the weak scale yet lower than
a heavy Higgs mass have been studied, see \cite{alfie}} behavior of cross
sections in the Standard Model has produced a series of surprising results
in recent years. In first place, this behavior is related to
the infrared structure of the theory, and not to the ultraviolet one as one
might na\"ively assume: this is due to the fact that in the electroweak
sector the symmetry breaking scale $M\sim 100$ GeV acts as an infrared
cutoff producing one loop radiative corrections that grow with the
c.m. energy $E$ like $\log^2\frac{E}{M}$ \cite{CC}. In second place, and related to
this, EW radiative corrections can become huge at the TeV scale, of the
order of $50\%$ for some LHC and ILC 
processes \cite{LHC}, opening the way for the
need to consider higher orders and resummation of leading effects. But the
greatest surprise comes out when one tries to define "infrared free"
observables, that are not affected by the above mentioned double logs: this
turns out to be impossible. In fact, even if additional EW gauge bosons in
the final state are included (fully inclusive observables), the
cancellation between "real" and "virtual" contributions that happens in QED
and QCD is spoiled by the fact that the EW symmetry is broken \cite{CCC00}.

At this point, one might be worried by the impossibility of making perturbative
predictions in the presence of radiative corrections that reduce by half
the value of tree level cross sections at the TeV scale. However, our
studies of resummations of leading effects
show that the asymptotic behavior is theoretically well under control and
can be summarized  in two lines:
\begin{itemize}
\item
All exclusive cross sections tend to zero \cite{FadinCC}.
\item
All inclusive cross sections tend to a linear combination of hard cross
sections \cite{CCC00}.
\end{itemize}
Here an ``exclusive'' observable is the one usually considered in the
literature: a definite final state is defined (say, two jets) and further
emission of weak gauge bosons is prohibited.  In the ``inclusive'' case,
all possible emissions of weak gauge bosons in the final state is included.
Since hard cross sections do not feature large logarithms (they typically
correspond to tree level quantities evaluated at the relevant hard scale),
the asymptotic behavior is well under control (see also \cite{CCC01}).
The measured observables fall somewhere between the ``fully
exclusive'' and ``fully inclusive'' case  
 depending on experimental cuts; the importance of evaluating gauge bosons emissions
for LHC observables has been emphasized in  \cite{Baur:2006sn}.

All the above holds in the ``recovered $SU(2) \otimes U(1)$'' limit: at
energies much higher than the weak scale, the leading interactions  fully respects
gauge symmetry and amplitudes and cross sections obey relations dictated by
the gauge symmetry: total weak isospin as well as total hypercharge are zero if
we consider momenta to be all incoming.  The purpose of this work is to
consider amplitudes that violate these quantum numbers, and therefore
vanish in the limit of unbroken gauge symmetry. 
On general grounds
these amplitudes must be zero when the vacuum expectation value (and
therefore the particles masses) go to zero,  and are 
suppressed by powers of $m/E$, $m$ being the relevant particle
mass, for very high energies $E$. Therefore one might think that these
amplitudes are negligible in the high energy limit; however, as we shall see,
the dressing by soft gauge bosons can lead to surprising results.

The basic model we consider contains two chiral spontaneously broken
gauge groups $U'(1)\otimes U(1)$. We  assume a large mass splitting
($M\gg m_{Z} $, $M$ being the Z' mass), so that 
the ${Z'}$-boson does not participate to the
IR dynamics. 
Thus, the $U'(1)$ allows
to construct simple amplitudes with total gauge charge violation, in our
case induced by the operator $\;\bar\psi\,{\cal
Z}_{\mu\nu}'\sigma^{\mu\nu}\,\psi\;$ (${\cal Z}_{\mu\nu}'=\partial_{\mu}{Z}_{\nu}'- 
\partial_{\nu}{Z}'_{\mu}$) describing the magnetic dipole
moment of the $Z'$ gauge boson. More explicitly, since left and right
fermion $U(1)$ hypercharges need not  be the same, the amplitude connecting
the $Z'$ with a left fermion and a right antifermion violates $U(1)$
(hyper)charge conservation. 
We compute the all order ($\alpha \log^2\frac{M^2}{m_Z^2}$) double leading logs (DLL),
taking care of the leading mass suppressed  corrections of order
  ${\cal O}\left(\frac{m^2}{M^2}\right)$, $m$ being the fermion mass.
We find that, among the form factors describing the effective  couplings of the $Z'$ to the  two light fermions, only the magnetic  one can develop 
exponentially growing 
Sudakov like corrections.

\section{The model}
We start by writing the most general Lagrangian describing the
 gauge bosons-fermion interactions (we assume usual kinetic terms for the abelian gauge bosons): 

\be
\bar{\psi}_L(
{\partial\hspace{-4.9pt}{\scriptstyle /}}+
i \,g\,y_L\,{Z\hspace{-4.9pt}{\scriptstyle /}}+i \,g'\,f_L\,{Z\hspace{-4.9pt}{\scriptstyle /}}')\psi_L+
\bar{\psi}_R(
{\partial\hspace{-4.9pt}{\scriptstyle /}}+
i \,g\,y_R\,{Z\hspace{-4.9pt}{\scriptstyle /}}+i \,g'\,f_R\,{Z\hspace{-4.9pt}{\scriptstyle /}}')\psi_R
\ee
where $\psi_{L/R}=\frac{1\pm\;\g_5}{2}\psi$,
$f_{L/R}$ ($y_{L/R}$) are the U'(1) (U(1)) hypercharges for left/right fermions.

 To implement, in a natural way, the spontaneous breaking 
of the gauge groups $U'(1)\otimes U(1)$ we need at least two complex Higgs fields,
 one, let's call $\phi'=\frac{1}{\sqrt{2}}(h'+v'+i\;\varphi')$
 with 
 $v'$ the $vev$  breaking $U'(1)$  and another scalar field, $\phi=
\frac{1}{\sqrt{2}}(h+v+i\;\varphi)$ 
with $v$ 
involved into the breaking of $U(1)$.
 The hierarchy $M_{Z'}\gg m_Z$ implies necessarily $v'\gg v$.
 The fermionic mass $m$ being of order $m_Z$ will be induced by the Yukawa interaction $\;\;h_f\;\bar \psi_R\;\phi\;\psi_L+h.c.$  so that  $m=\frac{h_f}{\sqrt{2}}\;v$ and for charge conservation we need both $f_{\phi}=f_R-f_L=2\;f_A$  and
$y_{\phi}=y_R-y_L=2\;y_A$.

Note that if \underline{$f_A\neq 0$} also the scalar field $\phi$ will participate to the breaking of $U'(1)$ and it will induce mixing  between   the gauge bosons $Z- Z'$ and  the Goldstone modes 
   $ \varphi' -\varphi$.

In order to clarify 
 the above considerations we  write the Lagrangian for the scalar sector
\footnote{The scalar potentials $V(\phi)$ and ${\cal V}(\phi') $ are
 responsible for the generation of the spontaneous symmetry breaking scales $v$ and $v'$}
\be\label{pot}
\left|(\partial_{\mu}+i\, g'\;f_{\phi'}\;Z_\mu')\phi')\right|^2+
\left|(\partial_{\mu}+i\, g'\;f_{\phi}\;Z_\mu'+i\, g\;y_{\phi}\;Z_\mu)\phi)\right|^2+(h_f\;\phi\;\bar{\psi}_R\;
\psi_L+h.c.)+V(\phi)+{\cal V}(\phi')
\ee 
and also the gauge fixing  Lagrangian (we choose to work in Feynman Gauge): 
\be\label{GF}
-\frac{1}{2}\left(  \partial_{\mu} Z^{\mu}-g\;y_{\phi}\;v\;\varphi \right
)^2
-\frac{1}{2}\left(  \partial_{\mu} Z^{'\mu}-M\chi'\right)^2;\quad
\chi'\equiv\frac{g'}{M}( f_{\phi}\;v\;\varphi+f_{\phi'}\;v'\;\varphi')
\ee


Working in the limit $\frac{ m_{f,Z}}{M}\ll 1$ we prefer to  use the gauge eigenstate basis as propagating free fields  with the mass shifts  used as perturbations.
In this limit the propagating $Z'$ field has mass
 $M^2=g^{'\,2}\;\left( f_{\phi}^2\;v^2+ \;f_{\phi'}^{2}\;v^{'\,2}\right)$, the $Z$ field has mass $m_Z^2=
 g^2\;y_{\phi}^2\;v^2$  and the mixing $Z'-Z$ is induced by the mass insertion
$\delta M^2=g\;g'\;f_{\phi}\;y_{\phi}\;v^2$ (note that $\frac{\delta M^2}{M^2}\ll 1$).

\section{Form factors  for the vertex $Z'\rightarrow \bar f \;f$}

The amplitude for Z' decay $Z'_{\mu}(p_1+p_2)\to \bar f(p_2)
\;{f}(p_1)$ is given by 
$\eps_\mu(p_1+p_2) \bar{u}(p_1)\Gamma_\mu^{(Z')} v(p_2)$ where
$\eps_\mu(p)$ is the physical Z' polarization satisfying 
$\sum_a \eps_\mu^a \eps_\nu^a=-g_{\mu\nu}+\frac{p_\mu p_\nu}{p^2}$.
In order to compute the effect induced by the multi loops generated by 
integrating over soft $Z$-gauge bosons, we introduce 
the more general CP invariant vertex for  with $(p_1+p_2)^2=M^2,p_1^2=p_2^2=m^2$:
     \be\label{ff} \bar{u}(p_1)\Gamma_\mu^{(Z')} v(p_2)=
i\; g'\;\bar{u}(p_1)\left [\g_\mu(F_LP_L+F_RP_R)+ 
\frac{m \; (p_1-p_2)_\mu}{(p_1\cdot p_2)}\; F_M+
\frac{m\; (p_1+p_2)_\mu}{(p_1\cdot p_2)}\; F_P\g_5\right]v(p_2) \ee
where $P_{R,L}=\frac{1}{2}(1\pm\g_5)$  and $F_{M}$ is usually named {\it
  magnetic   form factor}.
We also introduce  $F_V=\frac{1}{2}(F_R+F_L)$ and 
$F_A=\frac{1}{2}(F_R-F_L)$, the same relationships hold also for the tree
level charges $f_i$  and $y_i$. 

Defining  $\rho=\frac{m^2}{p_1p_2}$, the amplitudes squared for the various
positive (+) and negative (-) helicity fermions, summed over the Z' polarizations, are given by: 

 \be\label{Mhc}
\frac{ |{\cal M}_{++}|^2}{4(p_1p_2)}=\left(F_V-F_A \sqrt{1-\rho}\right)^2\qquad
\frac{ |{\cal M}_{--}|^2}{4(p_1p_2)}=\left(F_V+F_A \sqrt{1-\rho}\right)^2 \ee
 \be\label{Mhf} \frac{|{\cal M}_{+-}|^2}{4(p_1p_2)}= \frac{|{\cal M}_{-+}|^2}{4(p_1p_2)}
=\rho\;\left[F_A^2+(F_V-F_M(1-\rho))^2\right]
 \ee

The corresponding widths can be obtained multiplying by
 the appropriate phase space factors. Notice that, since $(p_1+p_2)^\mu
 \eps_\mu(p_1+p_2)=0$, the form factor $F_P$ does not contribute to
 physical amplitudes.

In the next section we calculate the on shell one loop form factors in the limit $M\gg
m_Z,m$, 
 retaining only the  DLL
contributions. Since we want to calculate the decay rates and the cross sections up to ${\cal O}(\rho )$,
we need the values of  $F_M$ to order $\rho^0$  and of $F_{L,R}$ 
 to order $\rho^1$.

\subsection{Form factors at One loop }

\begin{figure}
\centering \includegraphics[width=17cm]{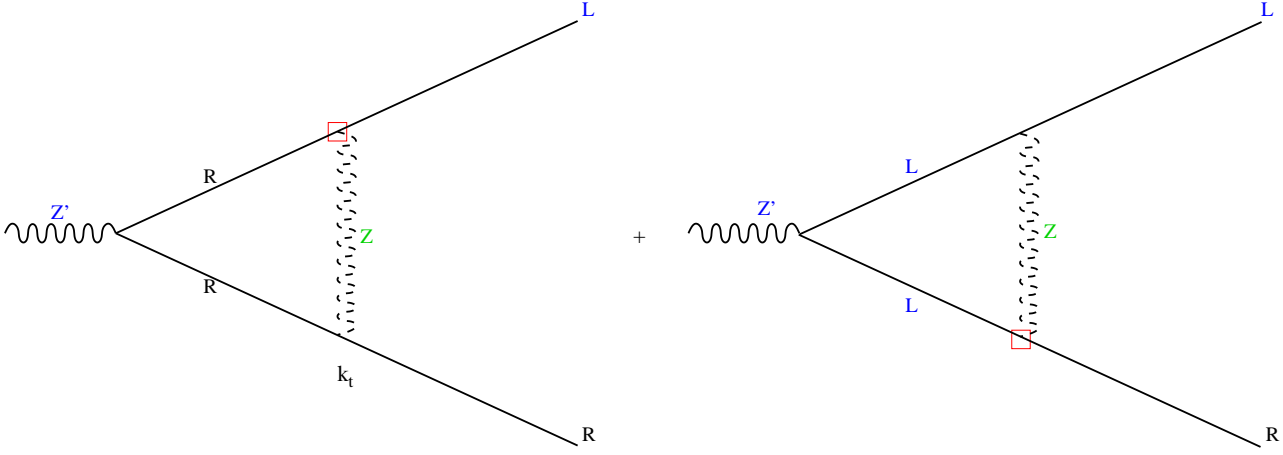}
\caption{\label{1loop} The anomalous magnetic moment at one loop. Diagram
(a) corresponds to a mass insertion ($\Delta_\mu$) on the upper (fermion) leg and an
eikonal current insertion ($J^{(eik)}_\mu$) on the lower (antifermion) leg (see eq.(\ref{correnti})).}
\end{figure}

Since we deal with on-shell external particles and physical  
polarizations, our amplitudes are gauge-invariant and can be computed  
in any gauge. Choosing to work in the Feynman gauge,
 we start analyzing carefully  the  one loop diagrams
depicted in   fig. (\ref{1loop}).
In the soft limit and at  DLL approximation,
we can neglect all terms proportional to the integration momentum $k$ in
the numerators, so the one loop contribution is:
  \ba g^2 \int \frac{d^4k}{(2 \pi)^4}\frac{-i}{k^2-m_Z^2} \!\!\!\!\!\!\!\!\!\!&&\bar{u}(p_1)\g_\alpha(y_LP_L+y_RP_R)
\frac{\psl_1+m}{2p_1k} \Gamma_\mu^0
\frac{\psl_2-m}{2p_2k}\g_\alpha(y_LP_L+y_RP_R) v(p_2),
\\ && {\rm with}\;\;\;
\Gamma_\mu^0=i\;g'\;\g_\mu(f_LP_L+f_RP_R) \nonumber
\ea
and, after some basic Dirac algebra, the one loop translates in this expression capturing all the double logs 
$\log^2\frac{M^2}{m_Z^2}$ :
\be \label{eq1l}
-\left(  \bar{u}(p_1) 
\;J_{1}^\alpha \;\;\Gamma_\mu^0
\;\;J_{2\alpha} \;v(p_2) \right)\;
\left(\int \frac{d^4k}{(2 \pi)^4}\frac{i}{k^2-m_Z^2} \;  \frac{ p_1p_2}{(p_1k)(p_2k)} \; 
\right)
\ee 
with  ( $y_A=\frac{y_R-y_L}{2}$) 
\be\label{corrent}
J_{2}^\mu=\frac{g}{\sqrt{p_1p_2}}
\left[p_{2}^\mu(y_LP_L+y_RP_R)-my_A\g^\mu\g_5\right] \qquad
J_{1}^\mu=\frac{g}{\sqrt{p_1p_2}}
\left[p_{1}^\mu(y_LP_R+y_RP_L)+my_A\g^\mu\g_5\right] 
\ee 
The one loop integral in double log  approximation gives (recall
that $M^2=(p_1+p_2)^2$):
\be
g^2\int \frac{d^4 k}{(2 \pi)^4}\frac{i}{k^2-m_Z^2}\frac{p_1p_2}{(p_1\cdot k)(p_2\cdot k)}=
g^2\int_{m_Z^2}^{M^2}
\frac{d \kt^2}{8 \pi^2}\frac{1}{\kt^2}\log\frac{p_1p_2}{\kt^2}
=\;\frac{\alpha}{4 \pi}\log^2\frac{M^2}{m_Z^2}\equiv L^2 ;
\qquad\alpha=\frac{g^2}{4\pi}
\ee
where subleading single log terms have been neglected. Projecting on the different Lorentz structures,
one obtains: \ba\label{magn}
 F_L^{(1)}&=& \left(-f_L\;y_L^2+\frac{\rho}{2}\;f_R\;(y_R^2-y_L^2)\right)\;L^2\qquad
 F_R^{(1)}= \left(-f_R\;y_R^2-\frac{\rho}{2}\;f_L\;(y_L^2-y_R^2)\right) \;L^2 \\
 F_M^{(1)}&=&y_A\;(f_L\;y_L-f_R\;y_R)\;L^2 \qquad
 F_P^{(1)}=y_A\;(f_L\;y_L+f_R\;y_R)\;L^2 \label{1pm}
 \ea

One can see  that the IR double logs affect both  ${\cal O}(\rho^0) $  and
 ${\cal O}(\rho)  $ corrections; the latter
 are proportional to the  $y_A$ charge of the fermions,  that is  non zero only 
 for chiral $U(1)$  gauge theories (clearly such double logs are not   presents in QED \cite{Ermolaev:2000sj}
and QCD).

It is instructive and useful for the next section  to separate the currents
 $J^i_\alpha$ defined
in (\ref{corrent}) in two components, the first one being the usual
eikonal current ($J^\mu_{\rm eik} $) and the second one ($\Delta^\mu$ ) 
responsible  of the chirality flip,
proportional to the fermion mass $m$ insertion 
\be\label{correnti} J_2^\mu=J^\mu_{\rm eik}(p_2)-\Delta^\mu, \qquad
J_1^\mu=J^\mu_{\rm eik}(p_1)+\Delta^\mu,
\;\qquad \Delta^\mu\equiv \frac{g}{\sqrt{p_1p_2}}\;m\;y_A\;\g^\mu\,\g_5 \ee 

\begin{figure}
\centering \includegraphics[width=15cm]{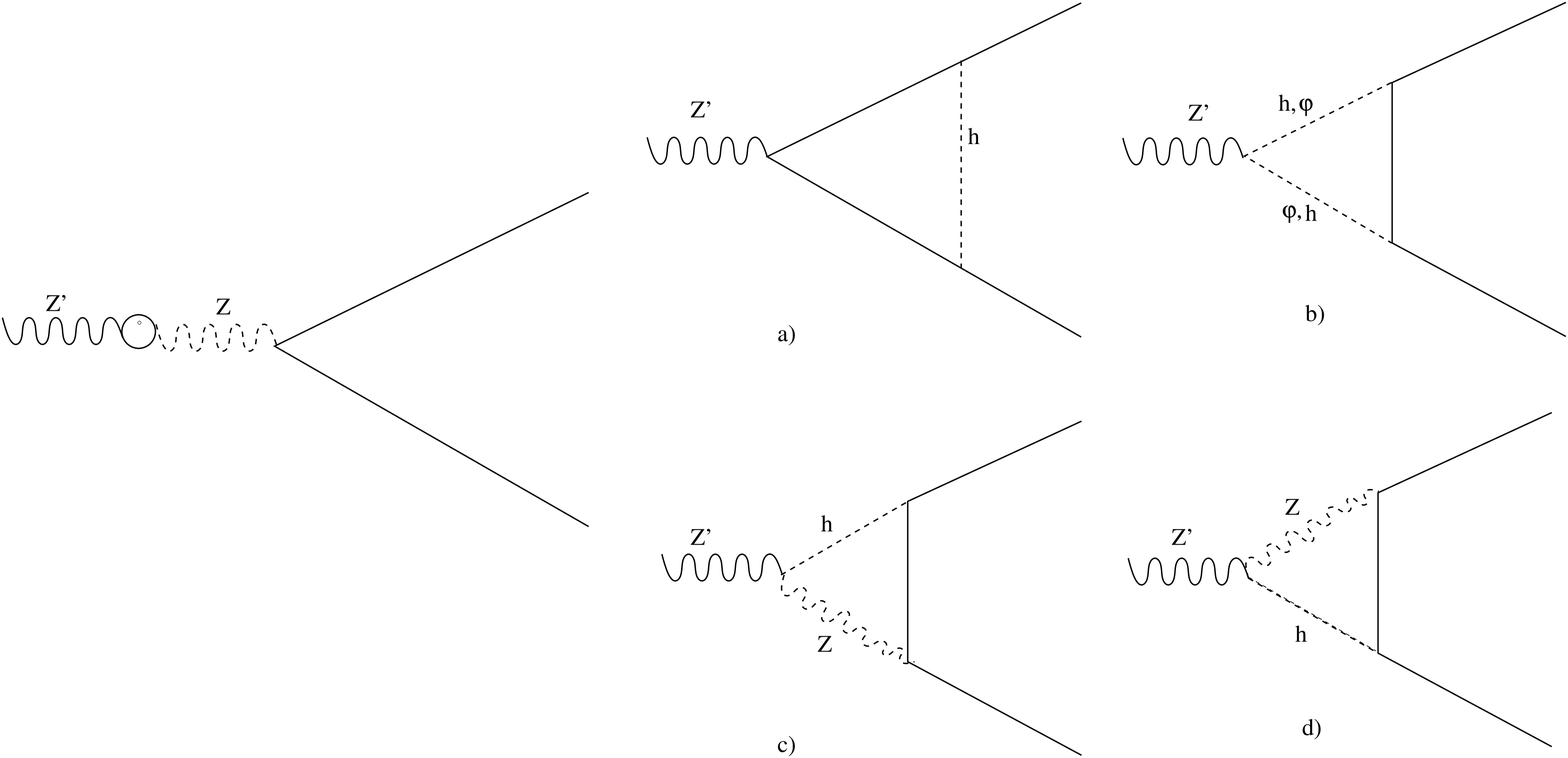}
\caption{\label{All}   Mixing and scalar loop effects to $Z'\rightarrow \bar{\psi}\psi$.
To the left we show the $Z'-Z$ mixing corrections.
 To the right we have the one loop scalar corrections  :
straight dotted line are the light scalars $h$ and $\phi$ while wavy lines are light $Z$ gauge bosons. }
\end{figure}

Let us now discuss another class of diagrams coming from $Z-Z'$ mixing and
from the Higgs/Goldstone sector (see fig. (\ref{All})).



\begin{itemize}
\item
Mixing effects in $Z'-Z$  sector 
(see the diagram to the left of fig.(\ref{All}))
simply produce a  shift $f_{L,R}\rightarrow f^0_{L,R}=f_{L,R}+
\frac{g}{g'}\;\frac{\delta M^2}{M^2}\;y_{L,R}$ 
so that eqs. (\ref{magn}, \ref{1pm}) are still valid with the replacement 
$f_{L,R}\to f^0_{L,R}$; the same holds for the all order resummed
expressions discussed in next section. In other words, $Z-Z'$ mixing only 
induces a (small ${\cal O}(\frac{m_Z^2}{M^2})$) renormalization of the
U'(1) fermion hypercharges.
\item The evaluation  of diagram a) gives 
\footnote{Note that only the scalar $h$, and  not the pseudoscalar $\varphi$,
contributes
at the DLL level.}
\be 
F_L^{(a)}= \frac{h_f^2}{16 \pi^2}\;\rho\;f_L\;\log^2\frac{M^2}{m_h^2}
\qquad
F_R^{(a)}= \frac{h_f^2}{16 \pi^2}\;\rho\;f_R\;\log^2\frac{M^2}{m_h^2}\qquad
F_{M,P}^{(a)}=0.
\ee 
 
\item The corrections from the  diagram $b)$ include 
two different contributions 
where the role of $h$ and $\varphi$ are interchanged; these two diagrams
have opposite signs and cancel out completely at the DLL level.
\item From diagrams c)+d) we get 
\be 
F_L^{(c+d)}= -\frac{\alpha}{\pi}\;\rho\;f_{\phi}\;y_{\phi}\;y_L\;log^2\frac{M^2}{m_{Z,h}^2}
\qquad
F_{R}^{(c+d)}=  -\frac{\alpha}{ \pi}\;\rho\;f_{\phi}\;y_{\phi}\;y_R\;log^2\frac{M^2}{m_{Z,h}^2}
\qquad
F_{M,P}^{(c+d)}=0.
\ee
 the   IR cutoff $m_{Z,h}$  is a mixing of the gauge and Higgs masses depending on their  relative magnitude (see \cite{long}).
\end{itemize}


\section{All order resummed form factors}

Since we  work in the regime $M\gg m_Z$,  our first order
 calculations cannot be trusted because $L^2\gg 1$, 
and we have to proceed to the resummation of all the DLL
($L^{2 n}$)  .  

The dressing by soft boson insertions of the eikonal type can be explicitly
taken into account at all orders by making use of the eikonal identity (see
\cite{Landau} for instance). We illustrate this calculation by adopting the
method of $\kt$-ordering: the leading terms in the resummed
series are given by ``ladder'' insertions ordered in the soft variable
$\kt$, which is the transverse momentum of the soft gauge boson\footnote{we
have  checked that the explicit computation using the eikonal identity
produces the same results obtained by  $\kt$-ordering}.

The resummation of the soft gauge bosons for momenta in the range 
$\kt^{inf}\leq\kt\leq \kt^{sup}$  is given by the following
 Sudakov form factor \cite{erm}:
\be\label{sud}
S_{i,j}[\kt^{sup},\kt^{inf}]=
\exp\left[-\frac{\alpha}{2\pi}y_i\,y_j
\int_{\kt^{2\,inf}}^{\kt^{2\,sup}}\frac{d\kt^2}{\kt^2}\log\frac{M^2}{\kt^2}\right]=
\exp\left[-\frac{\alpha}{4\pi}y_i\,y_j\;(\log^2\frac{M^2}{\kt^{2\,inf}}-
\log^2\frac{M^2}{\kt^{2\,sup}})\right]
\ee 
where  $y_i,y_j$ are the relevant $U(1)$ charges.

In general terms we have only three possible Sudakov structures: $S_{LL},\;S_{RR}$ and $S_{L,R}$ that after momenta integration will generate three kind of Sudakov exponents:
$e^{-y_L^2\;L^2},\;e^{-y_R^2\;L^2}$ and $e^{-y_L\,y_R\;L^2}$.
While the first two are exponentially suppressing their multiplicative factors, the last one ($e^{-y_L\,y_R\;L^2}$), that we call  \underline{Anomalous Sudakov}, depending on the sign of the charges $y_{L,R}$ can generate exponential growing corrections (for $y_L\;y_R<0$).

In fig. (\ref{fig3}) we schematically show the dressing  
of a one loop result, with 
 multiple insertions of soft gauge bosons. In fig. (\ref{fig3}) $b$ the red
boxes represent the ``cloud'' of all-order resummed DLL soft gauge bosons,
while the  wavy line is an insertion with a chirality-flip vertex ($\Delta_{\nu}$) on
the upper fermion leg and an eikonal insertion ($J_{eik}^{\nu}$) on the lower (antifermion) leg. 
The red box to the right represents ``very soft'' gauge bosons with 
transverse momenta $\kt'$ satisfying
$m_Z<\kt'<\kt$, $\kt$ being the wavy line $Z$ momentum;
 bosons in the red box to the left satisfy
$\kt<\kt'<M$. 

In order to get the all order  DLL amplitudes we 
show how the various   one loop terms are dressed by
the all order corrections:
\begin{itemize}
\item The double insertion of the two eikonal currents at one loop is replaced by the usual DLL Sudakov Form Factors for $F_{L,R}$:
\ba
-J^{\rm eik}_\alpha \;\Gamma_\mu^0 \;J^{\rm eik}_\alpha
\;\int_{m_Z^2}^{M^2}\frac{d \kt^2}{8 \pi^2}\frac{1}{\kt^2}
\log\frac{M^2}{\kt^2}
\rightarrow 
\;\g_{\mu}(S_{RR}[M,m_Z]\;P_R+S_{LL}[M,m_Z]\;P_L)
\ea

and the corresponding dressed form factors are
\ba\label{jj}
{ F}_L^{(J^2)}(s)=f_L\; S_{LL}[M,m_Z]=
f_L\;e^{- y_L^2L^2}\qquad
{ F}_R^{(J^2)}(s)=f_R\; S_{RR}[M,m_Z]=
f_R\;e^{- y_R^2L^2}
\ea
These terms represent  the "usual" Sudakov corrections coming from the DLL interactions that do not break the gauge symmetries.

\item 
The one loop diagram with one eikonal current on one leg and one 
chirality-flip current $\Delta_\mu$ on the other is dressed by soft 
$Z$ insertions (see fig.\ref{fig3}).
This  generates the ``anomalous''  $F_{M,P}$ form factors plus
an extra contribution to $F_{R,L}$ :

\begin{figure}
      \centering \includegraphics[height=5.5cm] {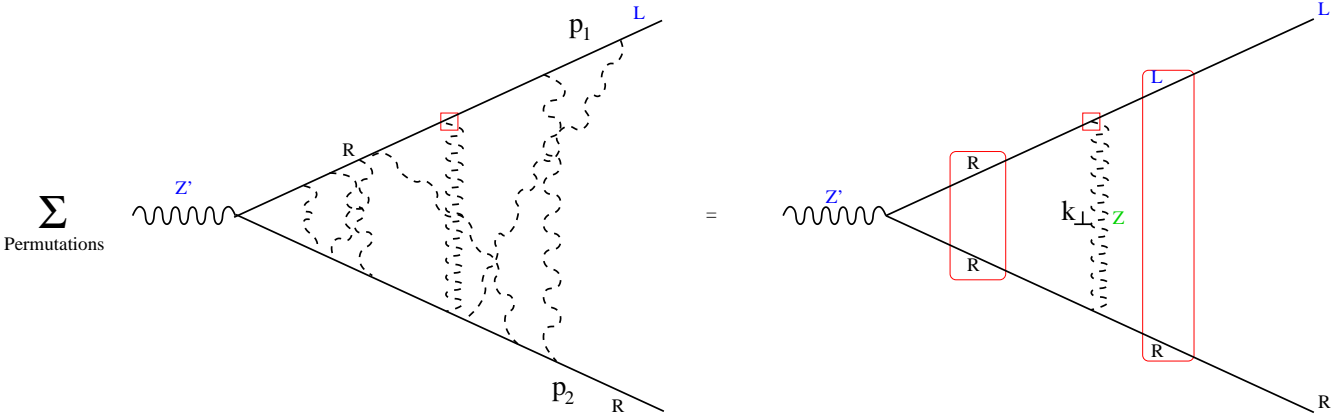}
      \caption{\label{fig3} All order Chirality breaking amplitude for the
process $Z'\rightarrow \bar f_R\;f_L$ mediated by the $Z$
bosons (dashed lines).  For simplicity the $Z'$ couples only to right
fermions.  To the left we draw the general structure of all the Feynman diagrams
that we have to sum up, to the right we used the ladder
diagram approach based on soft $\kt$ ordering where the red blocks means
Sudakov form factors (see eq. (\ref{sud})). }
         \end{figure}

\ba
\nonumber
\Delta_\alpha\Gamma_\mu^0J^{\rm eik}_\alpha
\;\int
\frac{d \kt^2}{8 \pi^2}\frac{1}{\kt^2}\log\frac{M^2}{\kt^2}
\rightarrow 
g^2y_A\frac{m\psl_2}{(p_1p_2)}\g_\mu
\int
\frac{d \kt^2}{8 \pi^2}\frac{1}{\kt^2}
\left(\!
-f_R\;y_R \; S_{RR}[M,\kt]\; \log\frac{M^2}{\kt^2}
 S_{LR}[\kt,m_Z]\;P_R \right.\\
\left.+  
f_L\;y_L \; S_{LL}[M,\kt]\; \log\frac{M^2}{\kt^2}
 S_{LR}[\kt,m_Z]\;P_L 
\right) 
\ea
\ba
\nonumber
J^{\rm eik}_\alpha\Gamma_\mu^0\Delta_\alpha
\;\int
\frac{d \kt^2}{8 \pi^2}\frac{1}{\kt^2}\log\frac{M^2}{\kt^2}
\rightarrow 
g^2y_A\g_{\mu}\frac{m\psl_1}{(p_1p_2)}\;\;\int
\frac{d \kt^2}{8 \pi^2}\frac{1}{\kt^2}
\left(
-f_Ry_R \; S_{RR}[M,\kt]\; \log\frac{M^2}{\kt^2}
 S_{LR}[\kt,m_Z]\;P_L
\right. \\ \left. +
f_Ly_L  \;S_{LL}[M,\kt]\; \log\frac{M^2}{\kt^2}
 S_{LR}[\kt,m_Z]\;P_R
\right)
\ea
The above contributions, sandwiched between the spinors $\bar{u}(p_1)$ 
and $v(p_2)$ produce the ``anomalous'' form factors
\ba\label{jd1}
F_M^{(J\Delta)}\equiv F_M=
\frac{1}{2}(f_L \;e^{- y_L^2 L^2}+f_R\; e^{-y_R^2 L^2})
-\frac{1}{2}(f_L+f_R)\;e^{-{ y_L\,y_R}\, L^2}
\\\nonumber
F_P^{(J\Delta)}\equiv F_P=
 \frac{1}{2}(f_L \;e^{-y_L^2 L^2}-f_R e^{- y_R^2 L^2 })
-\frac{(f_L-f_R)}{2}\;e^{- { y_L\,y_R}\, L^2}
\ea
and the corrections to $F_{L,R}$
\ba\label{cov1}
F_{L}^{(J\Delta)}= -\rho  \;f_R \left(
e^{- y_L y_R L^2}-e^{- y_R^2 L^2}\right)
\qquad 
F_{R}^{(J\Delta)}= -\rho \; f_L \left(e^{- y_L y_R L^2}-e^{- y_L^2 L^2}\right)
\ea
\item

 Finally the soft DLL gauge bosons dressing the one loop diagram with
 two $\Delta_\mu$ currents 
generates  a  contribution to $F_{R,L}$ (see fig.\ref{fig4}):
\ba 
\nonumber
- \Delta_{\alpha}\Gamma^0_{\mu}\Delta_\alpha
\;\int_{m_Z^2}^{M^2}\frac{d \kt^2}{8 \pi^2}\frac{1}{\kt^2}
\log\frac{M^2}{\kt^2}
\rightarrow 
 2m^2 \frac{g^2}{(p_1p_2)}y_A^2\;\g_\mu 
\int
\frac{d \kt^2}{8 \pi^2}\frac{1}{\kt^2}
\left(
f_L \; S_{LL}[M,\kt]S_{RR}[\kt,m_Z]\;P_R
\right.\\
\left. +
f_R  \;S_{RR}[M,\kt]S_{LL}[\kt,m_Z]\;P_L\;
\right)\; \log\frac{M^2}{\kt^2}
 \ea

\ba\label{dd}
\nonumber { F}_L^{(\Delta^2)}(s)&=&2
\frac{ \rho f_R y_A^2}
{
   \left(y_L^2-y_R^2\right)}
 \left(e^{- y_R^2L^2}-
e^{
   y_L^2 L^2}\right)
\qquad 
{ F}_R^{(\Delta^2)}=2
\frac{ \rho f_L y_A^2}
{\,
   \left(y_L^2-y_R^2\right)}
 \left(e^{- y_R^2L^2}-
e^{
   y_L^2L^2}\right)
\ea
\end{itemize}

\begin{figure}
      \centering \includegraphics[height=6cm] {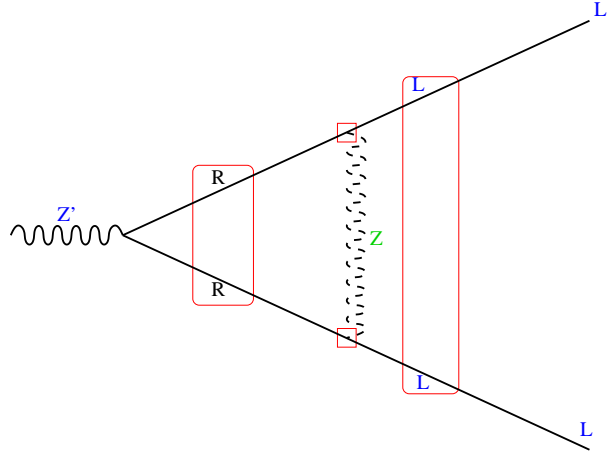}
      \caption{\label{fig4} All order Chirality breaking amplitude for the
process $Z'\rightarrow \bar f_L\;f_L$ mediated by $(\Delta_\mu)^2$ insertions. }
         \end{figure}

\begin{figure}
\centering \includegraphics[width=17cm]{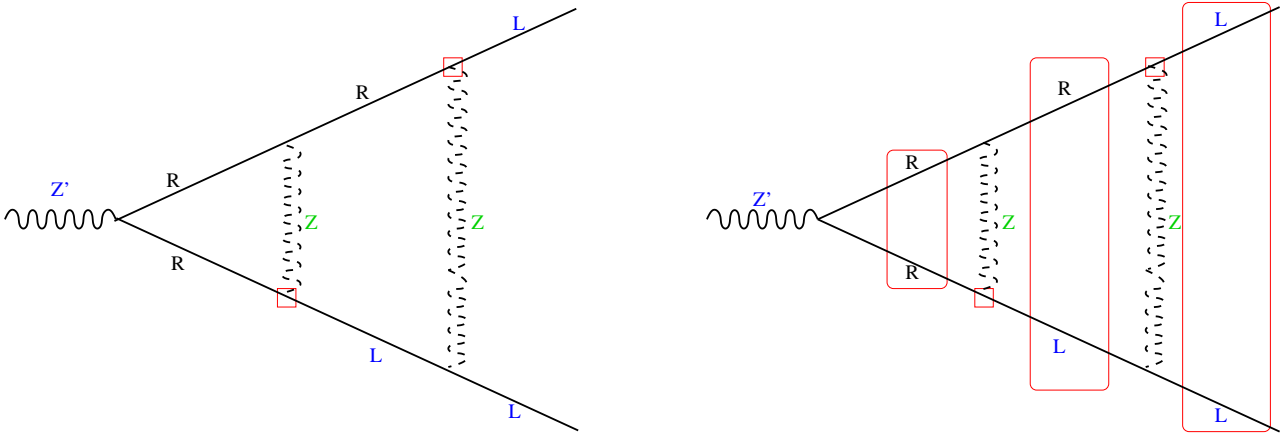}
\caption{\label{2loop}  The two loop (to the left) and the all order (to the right)  Chirality breaking amplitudes for the
process $Z'\rightarrow \bar f_L\;f_L$ mediated by  
$(\Delta_\mu\;J^{(eik)}_{\nu})^2$ 
insertions.}
\end{figure}

The resummed contribution to $F_{L,R}$ in eq. (\ref{cov1}) features an
``anomalous'' exponent proportional to $-y_Ly_R$, that can be positive for
left and right hypercharges of opposite signs. However this contribution,
as we show now, is canceled by a contribution from the DLL of the two
loop diagram in fig. \ref{2loop}, that  that also give  
${\cal O}(\rho)$ corrections to $F_{L,R}$.

The two loop expression is given by: 
\ba
2\;m^2\;\frac{g^2}{Q^4}\;y_A^2\;y_L\;y_R\;
\psl_2\g_\mu\psl_1
\left(
f_LP_R+f_RP_L
\right) 
\int_{m_Z^2}^{M^2} \frac{d \kt^2}{8\pi^2} \frac{1}{\kt^2}\; \log\frac{M^2}{\kt^2}\;
\int^{M^2}_{\kt^2} \frac{d \qt^2}{8\pi^2}
\frac{1}{\qt^2}
\; \log\frac{M^2}{\qt^2}
\ea 
while in order to evaluate the insertion of DLL soft gauge boson 
we have to evaluate the integral

\ba\nonumber
2\;m^2\;\frac{g^2}{Q^4}\;y_A^2\;y_L\;y_R\;
\psl_2\g_\mu\psl_1
\int_{m_Z^2}^{M^2} \frac{d \kt^2}{8\pi^2} \frac{1}{\kt^2}
\int^{M^2}_{\kt^2} \frac{d \qt^2}{8\pi^2}
\frac{1}{\qt^2}
\; \log\frac{M^2}{\kt^2}\; \log\frac{M^2}{\qt^2}
\left( f_L S_{LL}[M,\kt] S_{LR}[\kt,\qt] S_{RR}[\qt,m_Z]\;P_R+
\right.\\   \left. 
 f_RS_{RR}[M,\kt]S_{LR}[\kt,\qt]S_{LL}[\qt,m_Z]\;P_L \right)
\ea
The  result for the leading ${\cal O}(\rho)$ comes from
the substitution $\psl_2\ \g_\mu \ \psl_1\rightarrow -2 (p_1p_2) \g_\mu$ :
\ba\label{cov2}
F_{L}^{(J\Delta)^2}&=&  \;\rho\; f_R
\left(
e^{-y_Ly_RL^2}-\frac{1}{(y_L+y_R)}\left(
y_L\;e^{-y_R^2L^2}+y_R\; e^{-y_L^2L^2}\right)
\right)
\\\nonumber
F_{R}^{(J\Delta)^2}&=&  \;\rho\; f_L 
\left(
e^{-y_Ly_RL^2}-\frac{1}{(y_L+y_R)}
\left(
y_L\;e^{-y_R^2L^2}+y_R \;e^{-y_L^2L^2}
\right)\right)
\ea

Finally adding all together 
(eqs.(\ref{jj},\ref{jd1},\ref{cov1},\ref{cov2})) we obtain the   following results coming from pure $Z$ boson exchanges:
\ba\label{fulleq}
F_L ^{(Z)}&=& f_L\; e^{-y_L^2 L^2}-\frac{\rho}{2 }\;f_R
\;( e^{- y_R^2 L^2}- e^{- y_L^2 L^2})  \\ \label{fulleqR}
F_R ^{(Z)}&=& f_R \;e^{-y_R^2 L^2}-\frac{\rho}{2 }\;f_L\;
( e^{-y_L^2 L^2}- e^{-y_R^2 L^2})
\\
F_M^{(Z)}&=&
\frac{1}{2}(f_L \;e^{- y_L^2 L^2}+f_R\; e^{-y_R^2 L^2})
-\frac{1}{2}(f_L+f_R)\;e^{-{ y_L\,y_R}\, L^2}
\\
F_P^{(Z)}&=& \frac{1}{2}(f_L \;e^{-y_L^2 L^2}-f_R e^{- y_R^2 L^2 })
-\frac{1}{2}(f_L-f_R)\;e^{- { y_L\,y_R}\, L^2}
\ea

Let us now consider the all order dressing of the diagrams appearing in
fig. \ref{All}. As we are going to show now, these diagrams cannot produce
the ``anomalous'' effects we are studying here.

$\bullet$ The mixing effects in $Z'-Z$  sector  
simply amount to a renormalization
of the couplings; such renormalization is unphysical, as discussed previously

$\bullet$ The soft gauge boson cloud for the Higgs boson exchange 
of fig.(\ref{All}) , diagram $a)$,  gives (for $m_h=m_Z$):
\be \label{higgsoft}
F_L^{(a)}= \frac{h_f^2}{16 \pi^2}\;\rho\;f_L\;\log^2\frac{M^2}{m_Z^2}\;e^{-y_L^2\;L^2}
\qquad
F_R^{(a)}= \frac{h_f^2}{16 \pi^2}\;\rho\;f_R\;\log^2\frac{M^2}{m_Z^2}\;e^{-y_R^2\;L^2}
\ee

$\bullet $ For the scalar loop exchange we have to dress   only the diagrams c)+d) of fig.(\ref{All}).
In this case the dressing factor (\ref{sud})   is acting only on the external legs 
giving
\ba\label{hZ}
F_L^{(c+d)}\equiv F_L^{hZ}=-4\;g^2
\;f_{\phi}\;y_{\phi}\;y_L\;\rho\; \int
\frac{d \kt^2}{8 \pi^2}\frac{1}{\kt^2}log\frac{M^2}{\kt^2}
  S_{LL}[\kt,m_Z] =4\;f_{\phi}\;\frac{y_{\phi} }{y_L}\;\rho\; (e^{-y_L^2\;L^2}-1)
\\
F_{R}^{(c+d)}\equiv F_R^{hZ}= -4\;g^2
\;f_{\phi}\;y_{\phi}\;y_R\;\rho\; \int
\frac{d \kt^2}{8 \pi^2}\frac{1}{\kt^2}log\frac{M^2}{\kt^2}
  S_{RR}[\kt,m_Z] = 4\;f_{\phi}\;\frac{y_{\phi} }{y_R}\;\rho\; (e^{-y_R^2\;L^2}-1)
  \ea where we taken $m_h=m_Z$ for convenience.

Our conclusions about the terms generated by $Z'-Z$ mixing and the
Goldstone/Higgs sector are therefore the following:
\begin{itemize}
\item
Including $Z'-Z$ mixing only produces an unphysical (small)
renormalization of the couplings. 
\item
Terms produced by the Goldstone/Higgs
sector only affect the vector and axial form factors and are depressed  
at high energy by ``standard'' ($e^{-y_L^2\;L^2},\; e^{-y_R^2\;L^2}$)
Sudakov form factors
\item
All of these effects are unrelated to the ones produced by $Z$ exchange
since they are written in terms of independent parameters of the theory
$f_\phi,h_f,m_h$ and they vanish in some limit ($h_f\to 0, f_\phi\to
0$, heavy Higgs)
\end{itemize}
         
\vspace{0.2cm}

In order to obtain a cross check of our results we can
use Ward Identities (WI) that connect  the amplitude 
$\Gamma^\mu_{Z'\bar f f}$ with the amplitude 
$\Gamma_{\chi'\bar f f}$, $\chi'$ being the Goldstone boson of the $Z'$ (see eq.(\ref{GF})).
\be
\chi'=\frac{g'}{M}(f_{\phi'}\;v'\;\varphi'+f_{\phi}\;v\;\varphi)
\ee

The Goldstone $\chi'$ will interact with the fermions with coupling
\be\label{fa}
2\,f_{A}\;g'\; \frac{m}{M}\; \chi'\;\bar{u}(p_1)\;\g_5\;v_(p_2)
\qquad \underbrace{\rightarrow}_{\rm at\, all \; orders}\;\;\;\;
2\,F_{\chi'}\; g'\;\frac{m}{M}\; \chi'\;\bar{u}(p_1)\;\g_5\;v_(p_2)
\ee 
where $\,F_{\chi'}$ is the all orders form factor for the operator  $\chi'\;\bar\psi\;\g_5\;\psi$.

The relevant WI reads :
\be\label{wi}
(p_1+p_2)_\mu\;\Gamma^\mu_{Z'\bar f f}=
M\; \Gamma_{\chi'\bar f f}
\qquad{\rm or}\qquad
F_A+(1+\rho)\;F_P=\;F_{\chi'}
\ee
that at tree level ($F_A^{(0)}=f_A,\;\; F_P^{(0)}=0,\;\;F_{\chi'}^{(0)}=f_A$) is trivially satisfied.
 When we introduce the soft cloud ( at order ${\cal O}(\rho^0)$ ) we get the first non trivial consistency relation  after we explicit evaluate  $F_{\chi'}$ at all orders in $ L$ and at order
 ${\cal O}(\rho^0)$ (see fig. 5).  Evaluating the Feynman diagram   depicted in fig.5, with a 
 simple calculation we obtain
 
\begin{figure}
\centering \includegraphics[width=6cm]{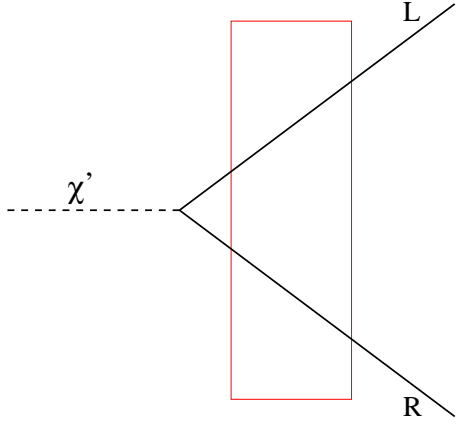}
\caption{\label{Fphi}   All order  DLL amplitude  for the Goldstone amplitude $\chi' \rightarrow \bar{\psi}\;\psi$. 
}
\end{figure}

\be
f_{\chi'}\rightarrow F_{\chi'}=f_{A}\;S_{LR}[M,m_Z]=f_A\;e^{-y_Ly_R\;L^2}
\ee

The check of eq.(\ref{wi}) at order ${\cal O}(\rho)$ requires a  calculation of  ${\cal O}(\rho^2)$
(due to the fact that we need  $F_P$ at order  ${\cal O}(\rho)$)
that is beyond the present  purposes.

\vskip.2cm

Overall, we can summarize the results of this section in the following way:
 \begin{itemize}
 \item the axial and vector form factors
related to $F_L$ and $F_R$ receive, after resummation, only ``standard'' Sudakov
form factors ($e^{-y_L^2\;L^2},\; e^{-y_R^2\;L^2}$) that exponentially suppress the amplitudes  at very large energies.
\item 
 The magnetic dipole moment form factor $F_M$ gets dressed also  with the  {\it Anomalous Sudakov} ($e^{-y_L\,y_R\;L^2}$)
 whose  exponent can  be {\sl positive} if $y_L\;y_R<0$.
 If this is the case, $F_M$  asymptotically dominates over $F_L,F_R$.
         \end{itemize}

\section{Asymptotic dynamics}

If $y_L\;y_R<0$, the terms proportional to the exponentially growing form
factor $F_M$ in the squared amplitudes (\ref{Mhc},\ref{Mhf})
dominate over the terms in $F_{L,R}$ for $M\gg m_z,m$. At what energy
scales $M$ does this happens ?

 Let us  consider, for simplicity, a vector like $Z'$ where $f_{L,R}\equiv f$ (in this  case  $f_A=0$ and all the  mixing terms  and scalar loops disappear). 
 The helicity changing
decay rate  $\Gamma_{+-}$ becomes:
\be
\Gamma_{+-}\simeq \Gamma_{+-}^{0}\;\;\frac{1}{4} \left(4\; e^{-2  y_L y_R L^2}+
e^{-2 y_R^2 L^2}-2\; e^{-
   \left(y_L^2+y_R^2\right)L^2}+e^{-2  y_L^2 L^2}\right)+{\cal O}(\rho^2)
\ee 
where  $\Gamma_{+-}^{0}$ is the tree level rate.
The resummed expression is a combination of
decreasing   and one potentially increasing (for $y_L\;y_R<0$) exponentials.
In the limit $L^2\gg 1$  and for  $y_L\;y_R<0$
quickly the
resummed value becomes twice as big as the tree level one, giving a 100 \%
radiative correction that puts in evidence the importance of the resummation.
 This happens
for scales such that:
\be \label{pheno}
e^{-2\;y_Ly_R{L}^2}=2\Rightarrow
\frac{M}{m_Z}=\exp[\sqrt{\frac{\pi\log 2}{-2y_Ly_R\alpha}}] \ee 
For
$y_L=-y_R=1,\alpha\sim1/30$ and $m_Z\sim$ 100 GeV
one obtains energies of the order of 30 TeV, which is a relatively low
scale value!
 
For  other observable  like the full decay rate $\bar\Gamma$ the expansion in $\rho$ gives (always taking $f_L=f_R=f$)
\ba\label{gamma}
\bar\Gamma\propto
f^2 (e^{-2  y_R^2 L^2}+ e^{-2  y_L^2 L^2})+
\rho\;f^2\;
\left(
2  e^{-2  y_L y_R L^2}+e^{-2  y_R^2 L^2}-2 e^{- \left(y_L^2+y_R^2\right)L^2}+e^{-2 
   y_L^2 L^2}
\right)+{\cal O}(\rho^2)
\ea
In this case the  anomalous Sudakov is always multiplied by a power of $\rho$.

If we compare  the $\rho=0$ terms with the anomalous exponential corrections, we see
 that they are of the same order when
\be
\rho \;e^{-2 y_L y_RL^2}\sim e^{-2  y_{R,L}^2 L^2}
\ee
and for $m\sim m_Z$ (just to have the order of magnitude) this happens at mass scales

%
\be
M\sim\; m\; e^{\frac{2\;\pi}{\alpha\;(y_{L,R}^2-y_Ly_R)}}
\ee
that  is of the same order  of  the Landau Pole (LP) energy
 $E_{LP}\sim \;m\; e^{\frac{\pi}{\beta\alpha}}$ ( where $\beta$  is the beta function of the $U(1)$ gauge group).
 
 \vspace{0.3cm}

Are these effects present  also into the SM?

It is  straightforward  to identify the chiral gauge group  $U(1)$ with $ U(1)_Y$
with $m_Z$  exactly the gauge boson $Z$ mass of $91$ GeV.
Then, from the analysis of the quantum number of the SM fields we see that
 $U(1)$ ``anomalous'' Sudakov form factors are presents only for the down quark sector where
$y_L=\frac{1}{6}$ and $y_R=-\frac{1}{3}$ so that $y_L\;y_R=-\frac{1}{18}<0$.

The phenomenological relevance of the above effects in this case result  quite suppressed  first of all for the smallness of the gauge  coupling $\alpha_Y\sim \frac{1}{60}$  and secondarily also for the 
smallness of the  charges $y_L\;y_R=-\frac{1}{18}$ .

The presence of anomalous Sudakov for the non abelian $SU(2)$ part is at present under study and results quite 
interesting because we naively expect  phenomenological relevant effects already at TeV scale  (see eq. (\ref{pheno}))
mainly due to the fact that the  gauge coupling is large ($\alpha_W\;\sim\; 2\,\alpha_Y$) and the  the  non abelian charges are naturally ${\cal O}(1)$ \cite{generic} .

\section{Conclusions}

In this work we have 
 evaluated the form factors of a very heavy $Z'$ gauge boson of mass $M$
into a
fermion-antifermion pair in a
simple $U(1)\otimes U'(1)$ model, performing the calculation up to
order $m^2$ in the fermion mass $m$ and to all orders in the $U(1)$  gauge coupling
at the double log level $(\alpha\;\log^2\frac{M^2}{m_Z^2})^n$.
We conclude that while the axial and
vector form factors feature a ``standard'', energy decreasing 
 Sudakov form factor, the magnetic dipole moment  features 
an ``anomalous'' 
 exponential $\sim \exp[{-\alpha\, y_L\, y_R\,\log^2\frac{M^2}{m_Z^2}}]$ term,
 which {\sl grows} with energy
  for fermions having opposite left-right  $U(1)$ charges ($y_L\;y_R<0$).  
This feature belongs exclusively to broken gauge theories like the
 electroweak sector of the Standard Model, and is a most unusual one. In
 fact the magnetic dipole moment corresponds to the insertion of an
 effective dimension five operator of the form  $\;\bar\psi_L\,{\cal
Z}_{\mu\nu}'\sigma^{\mu\nu}\,\psi_R\;$ (${\cal Z}_{\mu\nu}'=
\partial_{\mu}{Z}_{\nu}'- 
\partial_{\nu}{Z}'_{\mu}$), which explicitly breaks $U(1)$ if $y_L\neq y_R$
 and is (must be) proportional to the $U(1)$ vacuum expectation value. The
 expectation is that at large energy scales, where symmetry is recovered, 
 this symmetry violating operator  
gives negligible contribution to observables: this is by no means the
 case. While the contribution is truly suppressed by fermion masses at tree
 level, the dressing by IR dynamics around the light Z mass makes this
 operator the leading one at very high energies. This is a kind of ``non
 decoupling'' in the sense that very high energies observables are
 sensitive to the very low IR cutoff scale, whatever the ratio of the
 scales. This is due to the high energy behavior being dictated by the IR
 dynamics, and therefore sensitive to symmetry breaking at {\sl any} scale. 

  The main qualitative difference
 with respect to  all the previous Sudakov form factor evaluations 
(in QED and in QCD but  also in the high energy EW sector \cite{FadinCC}) 
 is the fact that the amplitudes we consider here do not  conserve 
 the gauge $U(1)$ charge of the soft $Z$ gauge bosons.
  In QED for photons and in QCD for gluons such conservation is automatic
 because the gauge symmetries are exact , while in the EW case up till now 
only  the leading  operators where no $SU(2)$ or $ U(1)_Y$ breaking  
is involved have been considered.
  The magnetic dipole moment instead is proportional to the chirality flip
 operator $\bar \psi_L\psi_R$ and from the $U(1)$ point of view carries a
 net $y_L-y_R$ charge:
this  is the main reason that allows for 
 the presence of unusual behavior in the  Sudakov form factors .
  As a result, the helicity flip width $\Gamma_{+-}$  can grow indefinitely at large energies.

A number of questions naturally arise: What is the role played by $Z$ emission
corrections? How the cancellation theorems \cite{KLN} involving real and virtual
corrections work? What happens
in the Standard Model itself, where the non abelian nature could
significantly
change things?  These problems will be addressed in future investigations.

\vspace{0.5cm}



{\bf Acknowledgments}

{\rm
D.C. during this work was  partially supported by the
EU FP6 Marie Curie Research \& Training Network "UniverseNet"
(MRTN-CT-2006-035863)}.


\def\np#1#2#3{{\sl Nucl.~Phys.\/}~{\bf B#1} {(#2) #3}} \def\spj#1#2#3{{\sl
Sov.~Phys.~JETP\/}~{\bf #1} {(#2) #3}} \def\plb#1#2#3{{\sl Phys.~Lett.\/}~{\bf
B#1} {(#2) #3}} \def\pl#1#2#3{{\sl Phys.~Lett.\/}~{\bf #1} {(#2) #3}}
\def\prd#1#2#3{{\sl Phys.~Rev.\/}~{\bf D#1} {(#2) #3}} \def\pr#1#2#3{{\sl
Phys.~Rep.\/}~{\bf #1} {(#2) #3}} \def\epjc#1#2#3{{\sl Eur.~Phys.~J.\/}~{\bf
C#1} {(#2) #3}} \def\ijmp#1#2#3{{\sl Int.~J.~Mod.~Phys.\/}~{\bf A#1} {(#2) #3}}
\def\ptps#1#2#3{{\sl Prog.~Theor.~Phys.~Suppl.\/}~{\bf #1} {(#2) #3}}
\def\npps#1#2#3{{\sl Nucl.~Phys.~Proc.~Suppl.\/}~{\bf #1} {(#2) #3}}
\def\sjnp#1#2#3{{\sl Sov.~J.~Nucl.~Phys.\/}~{\bf #1} {(#2) #3}}
\def\hepph#1{{\sl hep--ph}/{#1}}

\end{document}